\documentclass[twocolumn,
               prc,
               nofootinbib,
               showpacs,
	       floatfix]{revtex4}

\usepackage[T1]{fontenc}

\usepackage{amsmath}
\usepackage{graphicx}
\usepackage{color}
\usepackage{float}
\usepackage{dsfont}
\usepackage{xspace}

\pdfoutput=1
\usepackage{thumbpdf}
\usepackage[bookmarks]{hyperref}

\newcommand{\nnbbdec}{$0\nu2\beta$-decay }
\newcommand{\be}{\begin{equation}}
\newcommand{\ee}{\end{equation}}
\newcommand{\isotope}[2][]{{\ensuremath{{}^{#1}\mathrm{#2}}}} 
\newcommand{\kev}{\ensuremath{\mathrm{\,ke\kern -0.1em V}}\xspace}
\newcommand{\ev}{\ensuremath{\mathrm{\,e\kern -0.1em V}}\xspace}

\begin{document}

\title{Experimental study of double--$\beta$ decay modes using a CdZnTe detector array}

\author{J. V. Dawson$^{5}$, C. Goessling$^{1}$, B. Janutta$^{3}$, M. Junker$^{2}$, T. Koettig$^{1}$, D. Muenstermann$^{1}$, S. Rajek$^{1}$, C. Reeve$^{4}$, O. Schulz$^{1}$, J. R. Wilson$^{6}$, K. Zuber$^{3}$}
\affiliation{$^{1}$Lehrstuhl f\"ur Experimentelle Physik IV, Technische Universit\"at Dortmund, Otto--Hahn Str. 4, D--44227 Dortmund, Germany\\
$^{2}$Laboratori Nazionali del Gran Sasso, Assergi, Italy\\
$^{3}$Institut f\"ur Kern-- und Teilchenphysik, Technische Universit\"at Dresden, Zellescher Weg 19, D--01069 Dresden, Germany \\
$^{4}$University of Sussex, Falmer, Brighton BN1 9QH, UK\\
$^{5}$Laboratoire Astroparticule et Cosmologie, 10 rue Alice Domon et L\'eonie Duquet, 75205 Paris, France \\
$^{6}$University of Oxford, Denys Wilkinson Building, Keble Road, Oxford OX1 3RH,
United Kingdom}

\begin{abstract}
An array of  sixteen 1\,cm$^3$ CdZnTe semiconductor detectors was operated at the Gran Sasso Underground Laboratory (LNGS) to further investigate the feasibility of double--$\beta$ decay searches with such devices.  As one of
the double--$\beta$ decay experiments with the highest granularity the $4 \times 4 $ array accumulated an overall exposure of 18\,kg\,days. The setup and performance of the array is described.  Half-life limits for various double--$\beta$ decay modes of Cd, Zn and Te isotopes are obtained. No signal has been found, but several limits  beyond $10^{20}$ years have been performed. They are an order
of magnitude better than those obtained with this technology before and comparable to most other experimental approaches for the isotopes under investigation. An improved limit for the
$\beta^+$/EC decay of \isotope[120]{Te} is given.
\end{abstract}

\maketitle

\section{Introduction}

Following the past 20 years vast progress has been made in unveiling the
properties of the neutrinos. For decades neutrinos were thought to be massless, which no longer holds true: flavor oscillations found in the leptonic sector, studying neutrinos coming from the sun, the atmosphere, high energy
accelerators beams and nuclear power plants, are explained by a nonzero neutrino mass \cite{:2008zn,Shiozawa:2006qh,Ahmad:2002jz,Aharmim:2005gt,Robertson:2008zz,Abe:2008ee,Ahn:2006zza,Adamson:2008zt}.
However no absolute mass scale can be fixed with experiments studying the oscillatory behaviour. To achieve this, one has to investigate on weak decays, such as beta decay or neutrinoless double--$\beta$ decay ($0\nu2\beta$). The signature of the latter total lepton number violating process is the emission of two electrons with a sum energy corresponding to the $Q$--value of the nuclear transition.

The \nnbbdec is the gold plated process to distinguish whether neutrinos are Majorana or  Dirac particles. Furthermore, a match of helicities of the intermediate neutrino states is necessary, done
in the easiest way by introducing a neutrino mass. This mass is linked with the experimental observable half-life via
\begin{equation}
  \label{eq:1}
 \left(T_{1/2}^{0 \nu}\right) = G^{0 \nu}(Q, Z) \left| M_{GT}^{0\nu} - M_{F}^{0\nu} \right|^2 \left(\frac{\left< m_{ee}\right>}{m_e}\right)^2 \, ,
\end{equation}
where $\left<m_{ee}\right>$ is the effective Majorana neutrino mass, given by $\left< m_{ee} \right> = \left| \sum_{i}U_{ei}^2m_i\right|$ and $U_{ei}$ as the corresponding matrix element in the
leptonic PMNS mixing matrix, $G^{0 \nu}(Q, Z) $ is a phase space factor and $M_{GT}^{0\nu} - M_{F}^{0\nu}$ describes the nuclear transition matrix element.
Various stringent bounds on the half-life of several isotopes have been set. In addition, a potential evidence has been claimed in the  \nnbbdec of  $^{76}$Ge with $T_{1/2}^{0 \nu} = (2.23 \pm 0.4) \times 10^{25} a$ at 90\,\% C.L. \cite{KlapdorKleingrothaus:2006ff}.  For a recent review on double--$\beta$ decay see \cite{Avignone:2007fu}.
The COBRA experiment \cite{zuber} uses CdZnTe (CZT) semiconductors to search for $0\nu2\beta$--decay. CZT contains nine double--$\beta$ emitters, five of which can decay via the emission of two electrons (in decreasing order of their $Q$--value
these are \isotope[116]{Cd} ,\isotope[130]{Te} ,\isotope[70]{Zn}, \isotope[128]{Te} and \isotope[114]{Cd}), four of them can decay via double electron capture (EC) (\isotope[108]{Cd}), a combination of EC and $\beta^+$ decay (\isotope[64]{Zn} and \isotope[120]{Te}) or double $\beta^+$--decay (\isotope[106]{Cd}). Evidently, every isotope which is able to decay with positron emission can also decay via double EC. The positron decays have a very nice potential signal of two (four) 511\kev annihilation gammas, but the phase space of the decay is reduced with respect to pure electron capture modes. There is a revived interest in these decay modes, as it  has been shown that $\beta^+$/EC modes have an enhanced sensitivity to $V+A$ interactions \cite{Hirsch:1994es} and there might be a resonant enhancement in neutrinoless double EC into an excited state of the daughter nucleus if this state is
degenerate with the ground state of the mother isotope \cite{Sujkowski:2003mb}. The most promising isotope for the \nnbbdec in COBRA is \isotope[116]{Cd}  with a $Q$--value of 2809 keV resulting in a peak position  beyond all gamma lines occurring from the natural decay chains of U and Th.
There are a number of different experimental approaches, also working on the aforesaid isotopes. The most advanced measurement for \isotope[130]{Te} is done by CUORICINO using cryogenic bolometers. The resulting half-life limit for the neutrinoless decay of  \isotope[130]{Te} is given as   $T_{1/2} > 3.0 \times 10^{24}$ yrs (90 \% C.L.) \cite{Arnaboldi:2008ds}. Best limit for  \isotope[116]{Cd} comes from enriched
CdWO$_4$ scintillators \cite{Danevich03}. These kind of crystals and also ZnWO$_4$ scintillators are used to set limits on the remaining Cd and Zn isotopes mentioned  \cite{Belli:2008kz,Belli:2008zza}.
Recently  a search for $\beta^+$EC modes and double EC modes of $^{120}$Te has been performed using high purity germanium detectors. Results can be found in \cite{Barabash:2008zz}. Most of the obtained limits
are in the region of $10^{18}-10^{21}$ years. Results based on the semiconductor approach, namely the usage of CdZnTe detectors, were done with single detectors only \cite{Kiel:2003sm}
and on a small array of four detectors \cite{Bloxham:2007aaa}. Here we report on results of a four times larger array consisting of a layer of $4 \times 4$ detectors.

\section{Experimental Setup}

The data analyzed were taken with an array of $4\times 4$ CdZnTe
crystals installed in the COBRA setup at Laboratori Nationali
del Gran Sasso (LNGS).

CoPlanar Grid (CPG)\cite{luke1995} type CdZnTe detectors were used.
The detector crystals, manufactured by eV Products \cite{eVProducts},
were of cubic shape, measuring $11 \times 11 \times 11$\,mm$^3$ with a total mass of 103.9\,g.
Detector anodes and cathode were
gold-coated, the detector surfaces were passivated with a
manufacturer-specific red lacquer to provide a highly resistive surface
layer and to maintain long term stability.

The array was embedded in a custom-made Delrin (POM) holder-structure, electrical contacting
was done using thin Kapton- (Polyimide) Flex-PCB's and a low activity copper
loaded glue which was developed in-house. The whole setup is surrounded by a 5\,cm
inner copper shield, followed by 20\,cm of lead. A copper shield against
electromagnetic interference (EMI) and a neutron absorber consisting of 7\,cm
borated polyethylene completed the experimental setup. An overburden of
1400\,m of rock at the LNGS location provided an excellent muon shield
of 3500 m water equivalent.

The electronic readout chain used was custom made specifically for the
experiment. It consists of a 16 channel preamplifier with integrated
anode-grid subtraction circuit, low-noise supplies for grid-bias and
high voltage and a VME-data acquisition system using custom built ADCs.

We optimised all parameters relevant for detector performance --- specifically cathode bias voltage,
grid bias voltage and CPG subtraction circuit weighting factor --- individually
for each crystal. The whole setup was calibrated regularly with \isotope[22]{Na}
(511 and 1274.5\kev), \isotope[57]{Co} (122.1\kev) and \isotope[228]{Th} (238.6, 2614.5\kev).
The \isotope[228]{Th} calibration is especially useful, since it provides several lines
reaching from 84.4\kev up to the highest naturally occurring gamma energy of 2614.5\kev.

During the measuring period of the experiment, most of the time 12 detectors were active, yielding a
total active mass of 77.9\,g.

An in-depth description of the experimental setup, readout chain and detector
calibration is given in \cite{Dawson:2009ni}.

\section{Data Acquisition}

\begin{table}
 \begin{tabular}{l|c|c|c}
 \hline
 Isotope and Decay 				& Energy\,(MeV) & FWHM\,(\%)	& Effic. $\epsilon$\,(\%) \\ \hline
\isotope[116]{Cd} to gs			& 2.809		& 3.5 $-$ 8.8  	& 61.0 \\ \hline
\isotope[130]{Te} to gs			& 2.529		& 3.6 $-$ 8.9 	& 65.4 \\ \hline
\isotope[130]{Te} to 536\,keV 	& 1.993		& 3.8 $-$ 9.1 	& 58.2 \\ \hline
\isotope[116]{Cd} to 1294\,keV	& 1.511		& 4.1 $-$ 9.4 	& 71.4 \\ \hline
\isotope[130]{Te} to 1122\,keV	& 1.407		& 4.2 $-$ 9.5 	& 55.7 \\ \hline
\isotope[116]{Cd} to 1757\,keV	& 1.048		& 4.7 $-$ 10.0	& 60.5 \\ \hline
\isotope[128]{Te} to gs	 		& 0.868		& 5.1 $-$ 10.4	& 92.1 \\ \hline
\isotope[116]{Cd} to 2027\,keV	& 0.782		& 5.3 $-$ 10.6	& 68.2 \\ \hline
\isotope[130]{Te} to 1794\,keV	& 0.735		& 5.5 $-$ 10.7	& 63.0 \\ \hline
\isotope[116]{Cd} to 2112\,keV	& 0.697		& 5.6 $-$ 10.9	& 77.2 \\ \hline
\isotope[116]{Cd} to 2225\,keV	& 0.584		& 6.2 $-$ 11.4	& 77.1 \\ \hline
\isotope[114]{Cd} to gs	 		& 0.536		& 6.5 $-$ 11.7	& 96.5 \\ \hline \hline
\end{tabular}
\caption{$0\nu\beta^-\beta^-$ decay candidates and their decay energy together with the range of  energy
resolutions for all detectors (given as FWHM) at the decay energy and the average efficiency to detect the whole decay energy in one detector.
Efficiencies vary typically within 1-2 \% of the given value due to the different energy resolutions.}
\label{tab:DecayProperties}
\end{table}

The data analyzed in this paper can be divided into four periods. Small changes were applied during data taking which discriminate the periods from each other. After the first period, the whole setup was moved to a new location in the LNGS underground laboratory, but besides that, no changes were applied. For the third period, new preamplifiers were installed. During the fourth period, the detector channel assignment was changed. During periods one through three up to 12 crystals of the array were operational, while for the last period, only ten crystals were operational.

\begin{figure}[tb]
 \centering
 \includegraphics[angle = 90,width=\columnwidth]{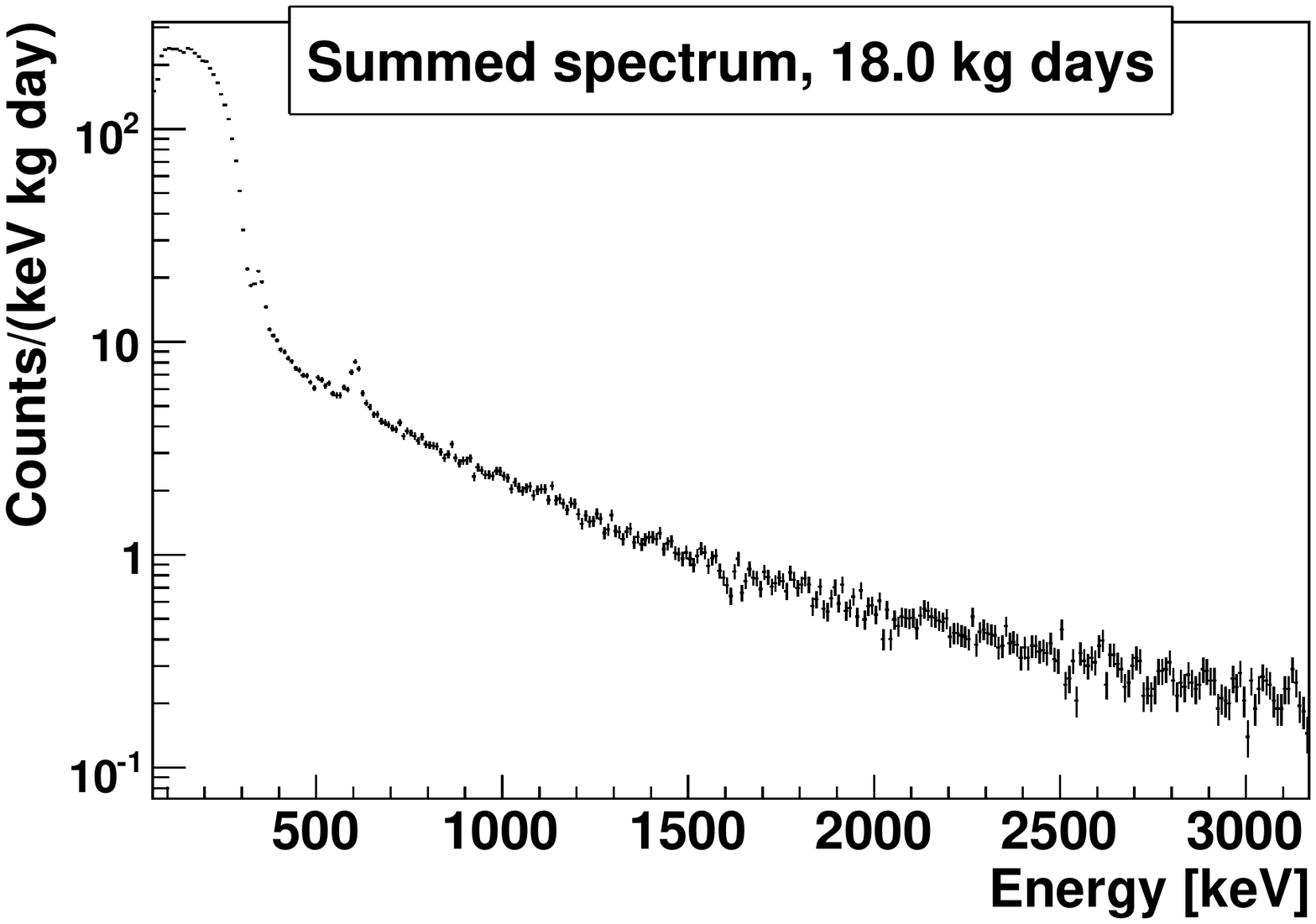}
 \caption{The overall sum spectrum acquired with 12 detectors. Clearly visible is a low energy shoulder at 300\kev due to the 4-fold forbidden non-unique beta decay of  \isotope[113]{Cd} . Furthermore,
 two $\gamma$--lines at 352\kev 609 \kev from the \isotope[238]{U} decay chain can be identified.}
 \label{fig:sumspectrum}
\end{figure}

Some cuts were applied to the data, to omit runs that have a high event rate due to e.g. microphonics resulting in piezo--electric discharges. A run is equivalent to an hour of data taking, counted for each detector separately. The first cut applied did reject files written by the data acquisition software with a size of more then 5\,MB (a typical run has about 200\,kB file size). In periods one through three this applied to 2634 out of 43580 runs. In the fourth period, 340 runs out of 25320 were rejected. Furthermore runs were rejected on a detector by detector basis. Therefore, a Poissonian was fitted to the number of counts above 500\kev and runs with a count rate exceeding the Poissonian with a probability higher than 99\,\% were rejected. Thus another 3157 runs out of 65926 remaining runs were cut (4.8\,\%). The spectrum of all usable runs is shown in \autoref{fig:sumspectrum}. No further events were removed from the raw data.

Variations observed in the resolutions could be attributed to the different preamplifiers used and the different voltages applied in the different data taking periods. It should be noted, that the detectors used during these measurements do not have the best energy resolution possible for CZT CPG semiconductor detectors. A better energy resolution would be desirable, but not essential at the moment, since the background was not known and therefore cheaper detectors were used.

\section{Data Analysis}

\begin{figure}[tb]
 \centering
 \includegraphics[width=\columnwidth]{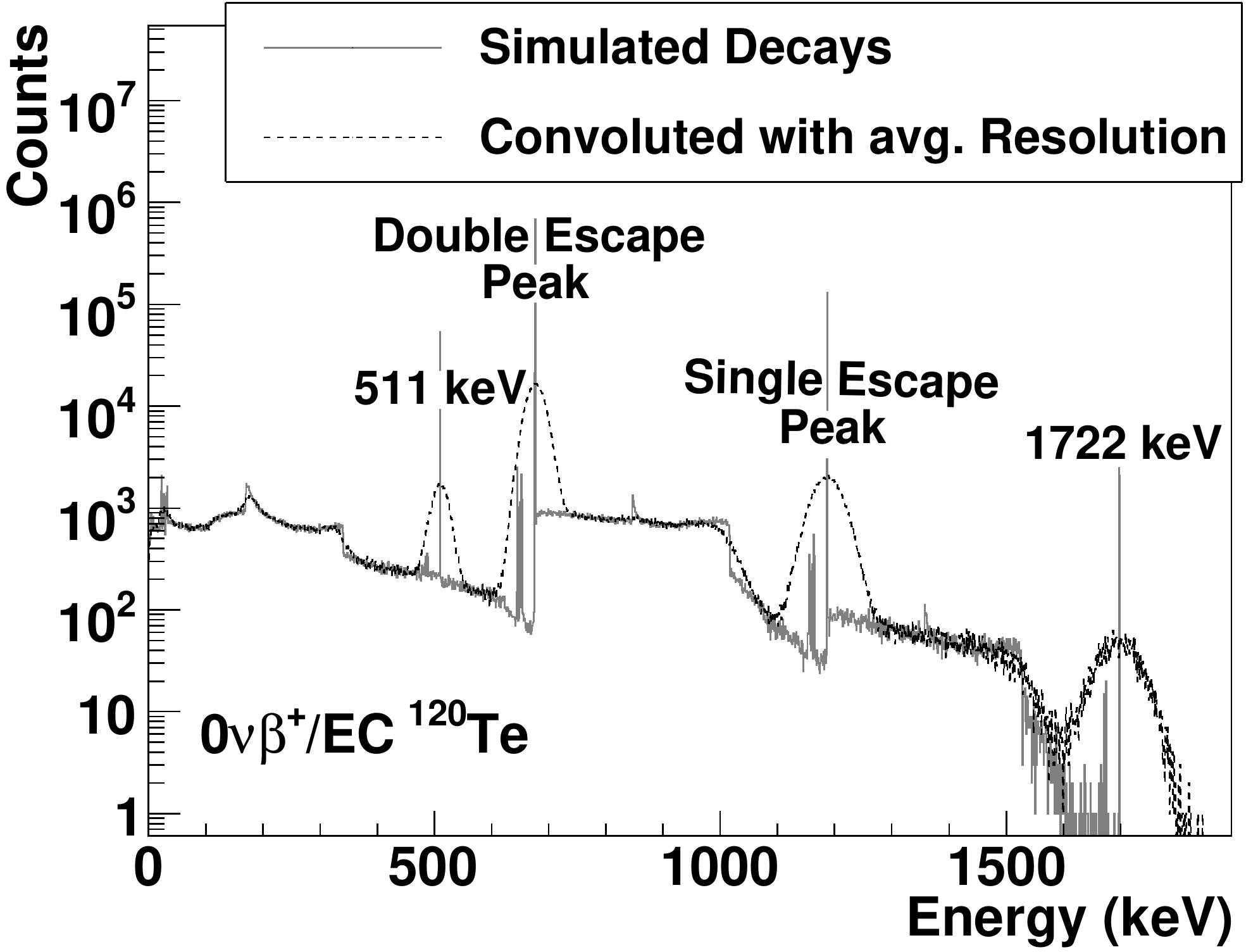}
 \caption{Simulated energy spectrum of  \isotope[120]{Te} $0\nu\beta^+$/EC to ground state as an example of the complexity to be expected in such decay modes. The decays were simulated for each detector individually assuming ideal resolution, but taking into account the setup geometry of the whole detector layer. Afterwards, the spectra were convoluted with the resolution of the given detector for each run individually. The spectrum is described in the text.}
 \label{fig:sim_b+b+}
\end{figure}
\begin{figure}[htb]
 \centering
 \includegraphics[width=\columnwidth]{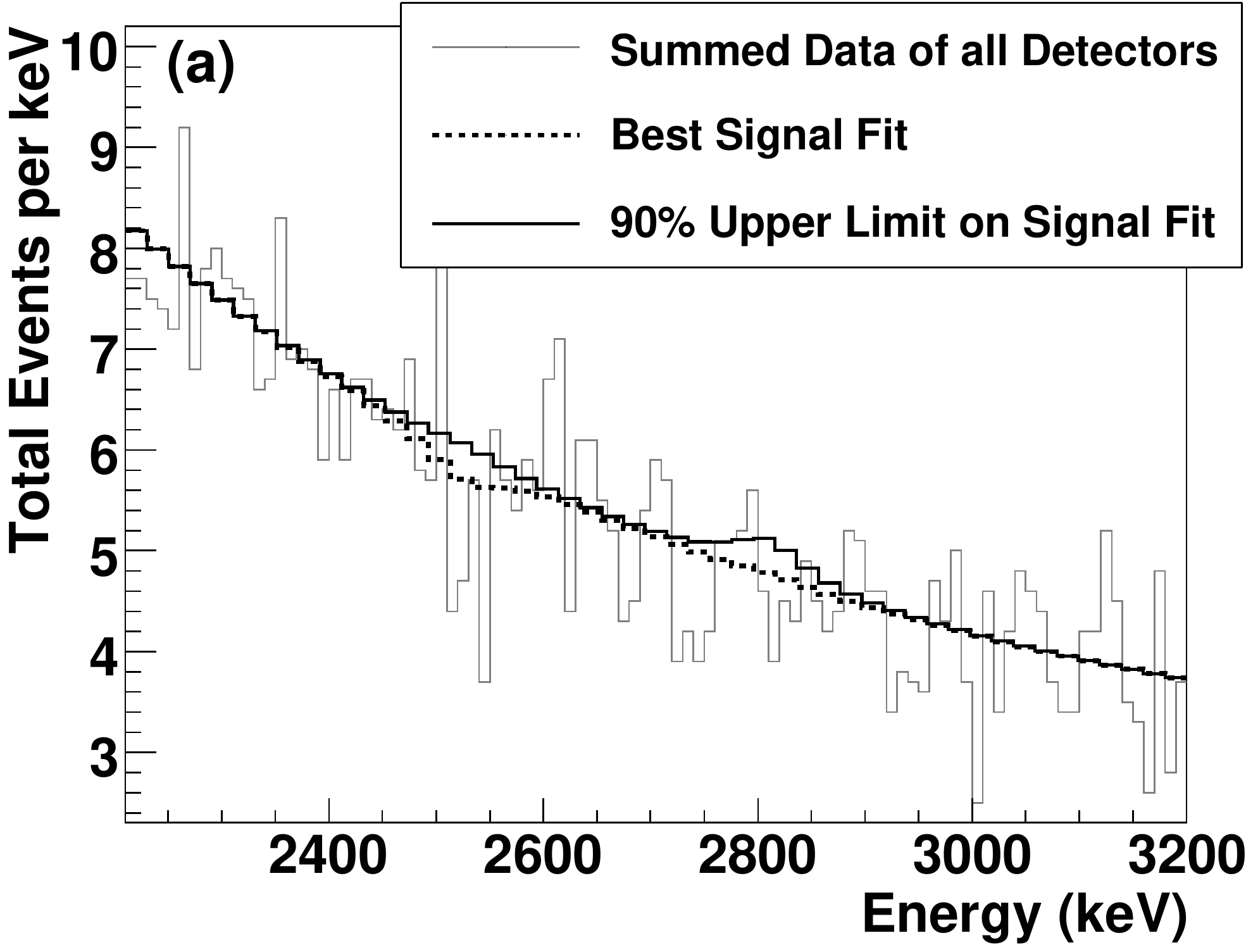}
 \includegraphics[width=\columnwidth]{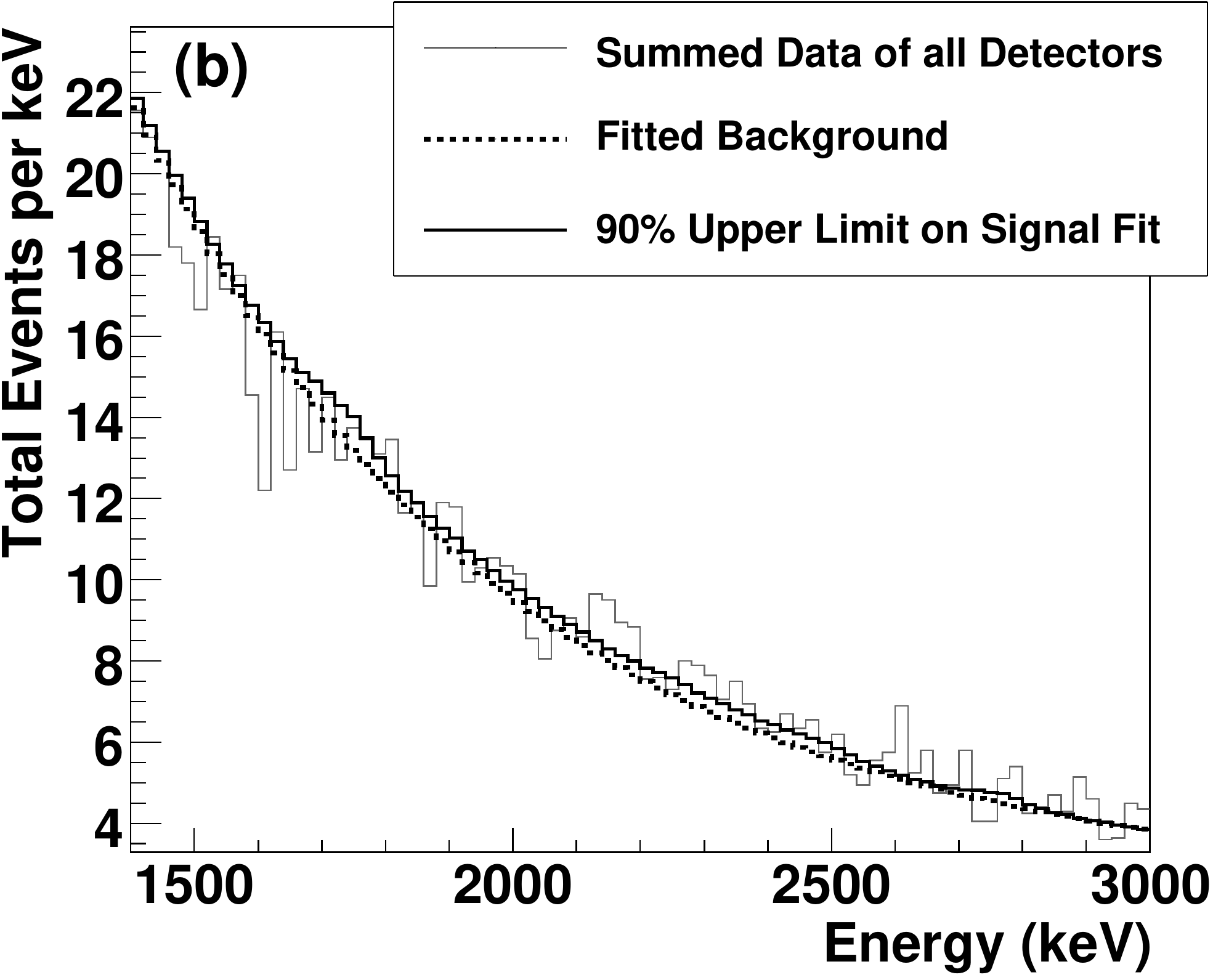}
 \caption{Fit results for \isotope[116]{Cd} (2809\,keV) and \isotope[130]{Te} (2529\,keV) $0\nu\beta^-\beta^-$  at Figure a) (combined fit for both isotopes), and \isotope[106]{Cd} EC/EC decays at Figure b). For the $\beta^-\beta^-$ isotopes, a Gaussian signal was fitted together with a background parametrization. For the $\beta^+\beta^+$ modes, simulated data sets were used to describe the signal.}
 \label{fig:0nubbfits}
\end{figure}
From the cleaned data sample of 18 kg\,days, upper limits on the number of signal events were extracted using a maximum likelihood fit. Therefore, the data was histogrammed for each detector. For the fit, each bin of a histogram was modeled as a Poisson random variable depending on the signal and background rate. The complete likelihood function, deriving from the Poissonian probability distribution, for all detectors and all runs during the measurements is given by
\begin{equation}
 \ln(L) = \sum\limits_{\rm Det.} \, \sum\limits_{\rm Run} \, \sum\limits_{\rm Bin} \left( D_{\rm Bin} \cdot  \ln F_{\rm Bin}) - F_{\rm Bin} \right)
\end{equation}
with the number $D_{\rm Bin}$ of observed events in the given histogram bin and the fitted number $F_{\rm Bin}$ of counts determined from the fit function. This approach takes into account the energy resolution and background level of each detector individually by the parameters of $F_{\rm Bin}$ and also allows the consideration of changes of resolution and background levels during different runs. The log-likelihood function was minimized with the MINUIT \cite{James:1975dr} package.\par
The fit model for $F_{\rm Bin}$ consists of a background parametrization together with a signal parametrization. No background was subtracted from the analyzed data sample. The background model contains two exponentials, one describing the higher energetic spectrum above 700\,keV, the other describing the spectrum above 400\,keV. Additionally, a Gaussian is fitted to the 609.3\,keV \isotope[214]{Bi} line. Below 400\,keV, the spectrum is mainly dominated by the fourfold forbidden $\beta$-decay of \isotope[113]{Cd} \cite{Dawson:2009ni} and low-energetic $\gamma$ lines like the 351.9\,keV \isotope[214]{Pb} line. As the lowest peak investigated is at 534\kev, data below 400\,keV were not taken into account for the analysis. \par
For the signal model, different approaches for $0\nu\beta^-\beta^-$ and $0\nu\beta^+\beta^+$ decays were taken. The summed energy of the two electrons of a $0\nu\beta^-\beta^-$ decay results in a Gaussian signal peak with a mean at the decay energy $E_{peak}$. Such a Gaussian was taken as signal model for the $\beta^-\beta^-$ decays. To ensure a proper fit of the background, a fit range of at least three peak widths (FWHM) around the expected peak energy was chosen. Due to the close proximity of the expected signal peaks of  \isotope[116]{Cd} and \isotope[130]{Te} as well as the \isotope[70]{Zn} and \isotope[128]{Te} ground state transitions, a combined fit of both peaks was made for these signals. To obtain the efficiency $\varepsilon$ for observing the full decay energy, Monte Carlo simulations taking into account the escape probabilities of the electrons and also of resulting gammas were done for each detector. The expected $0\nu\beta^+\beta^+$, namely double positron transition $\beta^+\beta^+$, single electron capture $\beta^+$/EC and double electron capture EC/EC spectra are more complex than the $\beta^-\beta^-$ ones. They normally contain several lines deriving e.g. from the escape of 511\,keV annihilation gammas from the detectors, but do not have a dominating peak. An example is given in \autoref{fig:sim_b+b+} with the simulation of $0\nu\beta^+$/EC decays of \isotope[120]{Te}. There, besides a peak at the full $Q$--value of 1722\,keV and the mentioned escape peaks, also satellite lines 22\kev below these lines deriving from additional escape of a $K_\alpha$ X-ray and a peak at 511\,keV deriving from gammas produced in annihilations of positrons from the $\beta^+$ decays in other detectors of the 16er layer are visible.\par
To cope with these complex signal distributions, an alternate approach for the search for these signals was taken. The spectra of the $0\nu\beta^+\beta^+$ decay modes were simulated for each detector taking into account the mass of each detector and its position inside the layer. The resulting spectrum was convolved with the energy resolution of the given detector, normalised to unity and taken as signal model. The fit range for every isotope includes the main lines of the given decay.\par
Results of the fit of the background and the signal model for the  $0\nu\beta^-\beta^-$ decays of \isotope[116]{Cd} and \isotope[130]{Te} and for the $\beta^+$/EC decay of \isotope[106]{Cd} are shown in \autoref{fig:0nubbfits}.\par
The simulation of $\beta^+\beta^+$ and $\beta^-\beta^-$ signals in the detectors of the experimental setup were done with a GEANT4 based Monte Carlo simulation. The Fortran Decay0 \cite{Decay0Article} code was used as event generator.\par
\begin{figure}[htb]
 \centering
 \includegraphics[width=\columnwidth]{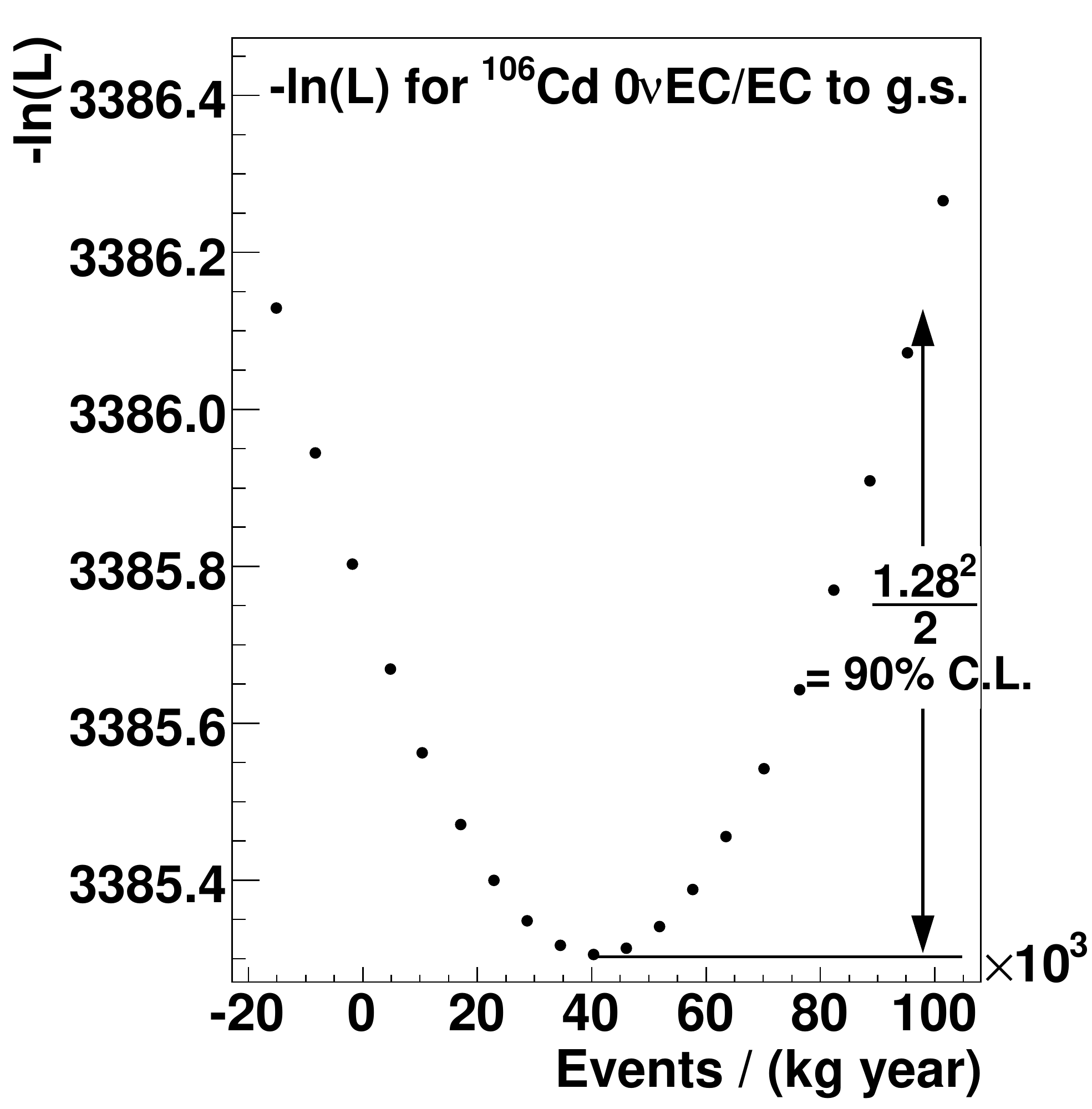}
 \caption{The shape of the negative log-likelihood function for the EC/EC fit of \isotope[106]{Cd}. The 90\% upper confidence level is determined by the point where $-\ln(L)$ rises $\frac{1.28^2}{2}$ over the minimal value.}
 \label{fig:scanlikelihoodfkt}
\end{figure}

A 90\% confidence level upper limit on the fitted counts was derived from the shape of the log-likelihood function. It is defined as the point where the (negative) log-likelihood function rises by a value of $\frac{1.28^2}{2}$ above its minimum. The shape of the log-likelihood function for the \isotope[106]{Cd} $0\nu$EC/EC fit is shown in \autoref{fig:scanlikelihoodfkt}.
In case the count rate obtained by the fit is positive, the 90\% C.L. upper limit $N_{\rm sig}$ on signal events is calculated by summing the fit result and the associated 90\% C.L. error. In case of it being negative, a conservative approach by only using the 90\% C.L. error has been applied. With the obtained upper limit, a 90\% lower half-life limit can be calculated via
\begin{equation}
 T_{1/2} > \frac{\ln(2) \cdot N_{\rm iso} \cdot t_{\rm live} \cdot \epsilon}{N_{\rm sig}} \quad ,
\end{equation}
where $N_{\rm iso}$ is the number of source atoms under study, $t_{\rm live}$ is the lifetime of the experiment and $\epsilon$ is the above mentioned simulated efficiency of observing a decay in the experimental setup as given in \autoref{tab:DecayProperties}. For the $\beta^+\beta^+$ decays, the whole simulated spectrum was used for the extraction of excluded events, thus $\epsilon = 1$ holds for these calculations.

The Cd and Zn content of the Cd$_{0.9}$Zn$_{0.1}$Te is only known to lie in a range of 7\% to 11\% for Zn and accordingly 89\% to 93\% for Cd. Therefore, conservatively a zinc content of 7\% was assumed when calculating the limits for Zn isotopes and 11\% zinc content was taken for the Cd isotopes.

\section{Results}

\begin{table}
 \begin{tabular}{|c|c|c|c|}
 \hline \hline
 Isotope and Decay 	& Fit Range 	& \multicolumn{2}{|c|}{$\mbox{T}_{1/2}$ limit (years)} \\ \cline{3-4}
			&  (MeV)	& This work	& Measured \\ \hline
\isotope[116]{Cd} to gs	& 2.2$-$3.2	& $9.4 \times 10^{19}$  & $1.7 \times 10^{23}$ \cite{Danevich03} \\
\isotope[130]{Te} to gs & 2.2$-$3.2	& $5.0 \times 10^{20}$  & $3.0 \times 10^{24}$ \cite{Arnaboldi08} \\ \hline
\isotope[130]{Te} to 536\,keV  & 1.7$-$2.3	 & $3.5 \times 10^{20}$& $9.7 \times 10^{22}$ \cite{Alessandrello00} \\ \hline
\isotope[116]{Cd} to 1294\,keV	 & 1.2$-$1.8	 & $5.0 \times 10^{19}$ & $2.9 \times 10^{22}$ \cite{Danevich03} \\ \hline
\isotope[116]{Cd} to 1757\,keV	 & 0.9$-$1.3	 & $4.2 \times 10^{19}$  & $1.4 \times 10^{22}$ \cite{Danevich03}\\ \hline
\isotope[128]{Te} to gs	 & 0.6$-$1.3	 & $1.7 \times 10^{20}$ 	 & $1.1 \times 10^{23}$ \cite{Arnaboldi05} \\ \hline
\isotope[116]{Cd} to 2027\,keV & 0.5$-$1.2	 & $2.8 \times 10^{19}$ 	 & $2.1 \times 10^{21}$ \cite{Piepke94}  \\ \hline
\isotope[116]{Cd} to 2112\,keV & 0.5$-$1.0	 & $4.7 \times 10^{19}$ 	 & $6.0 \times 10^{21}$  \cite{Danevich03} \\ \hline
\isotope[116]{Cd} to 2225\,keV & 0.5$-$1.0	 & $2.1 \times 10^{19}$ 	 & $1.0 \times 10^{20}\, ^\dagger$ \cite{Barabash90}  \\ \hline
\isotope[130]{Te} to 1794\,keV	 & 0.5$-$1.2	 & $1.9 \times 10^{20}$ 	 & $2.3 \times 10^{21}$ \cite{Barabash01}\\ \hline
\isotope[130]{Te} to 1122\,keV	 & 1.1$-$1.7	 & $1.2 \times 10^{20}$ 	 & $2.7 \times 10^{21}$ \cite{Barabash01} \\ \hline
\isotope[114]{Cd} to gs	 & 0.4$-$1.0	 & $2.0 \times 10^{20}$ & $1.1 \times 10^{21}$ \cite{Belli:2008zza}	  \\ \hline \hline
\end{tabular}
\caption{Results for fits of $0\nu\beta^-\beta^-$ decays on 90\% C.L. and fit ranges. Entries marked with $^\dagger$ are on 68\% C.L. Isotopes not separated by a horizontal line were fitted together.}
\label{tab:b-b-results}
\end{table}

\begin{table}
 \begin{tabular}{|c|c|c|c|}
 \hline \hline
 Isotope and Decay 	& Fit Range 	& \multicolumn{2}{|c|}{$\mbox{T}_{1/2}$ limit (years)} \\ \cline{3-4}
		&  (MeV)	& This work	& Measured \\ \hline
\isotope[64]{Zn} $\beta^+$EC to gs	 & 0.5$-$1.1	& $1.1 \times 10^{18}$ 	 & $4.3 \times 10^{20}$ \cite{Belli:2008kz} \\ \hline
\isotope[64]{Zn} 2EC to gs	 & 0.5$-$1.3	& $3.3 \times 10^{17}$ 	 & $1.1 \times 10^{20}$ \cite{Belli:2008kz} \\ \hline
\isotope[120]{Te} $\beta^+$EC to gs	 & 1.0$-$2.0	& $4.1 \times 10^{17}$  & $1.9 \times 10^{17}$ \cite{Barabash:2008zz} \\ \hline
\isotope[120]{Te} 2EC	 & 0.8$-$2.0	 & $2.4 \times 10^{16}$  & $6.0 \times 10^{17}$ \cite{Barabash:2008zz} \\ \hline
\isotope[120]{Te} 2EC to 1171\,keV	 & 0.6$-$2.0	 & $1.8 \times 10^{16}$	 &  $6.0 \times 10^{17}$ \cite{Barabash:2008zz} \\ \hline
\isotope[106]{Cd} $\beta^+\beta^+$ to gs. & 0.5$-$2.0	 & $2.7 \times 10^{18}$ 	 & $2.4 \times 10^{20}$ \cite{Belli99} \\ \hline
\isotope[106]{Cd} $\beta^+$EC to gs	 & 1.5$-$3.0	 & $4.7 \times 10^{18}$ 	 & $3.7 \times 10^{20}$ \cite{Belli99} \\ \hline
\isotope[106]{Cd} 2\,EC to gs	 & 2.0$-$3.0	 & $1.6 \times 10^{17}$ & $3.5 \times 10^{18}$ \cite{Barabash96} \\ \hline
\isotope[106]{Cd} $\beta^+ \beta^+$ to 512\,keV	 & 0.6$-$1.5	 & $9.4 \times 10^{17}$ 	 & $1.6 \times 10^{20}$ \cite{Belli99} \\ \hline
\isotope[106]{Cd} $\beta^+$EC to 512\,keV	 & 0.8$-$2.0	 & $4.6 \times 10^{18}$ & $2.6 \times 10^{20}$ \cite{Belli99} \\ \hline \hline
\end{tabular}
\caption{Results obtained for $0\nu\beta^+\beta^+$ modes on 90\% C.L. and fit range.}
\label{tab:b+b+results}
\end{table}

The obtained results for $\beta^-\beta^-$ decays are listed in \autoref{tab:b-b-results} and for $\beta^+\beta^+$ in \autoref{tab:b+b+results}. Due to the low source mass of $77.9$\,g not all results can compete with the results of world leading large scale experiments. However, former COBRA limits could be significantly improved, six of them by more than an order of magnitude and several have passed the $10^{20}$ yr boundary. In  addition, various decay channels are within an order of magnitude of  the world leading limits.

Due to significant improvements possible in the near future there is a good chance to surpass some of the existing limits. The installation of a nitrogen atmosphere including a radon trap combined with the replacement of the red passivation lacquer indicates a reduction of background by an order of magnitude. The installation of 48 more detectors will add another factor three in source strength and also significantly increase the
sensitivity for several decay modes especially those involving photons. Furthermore, a potential coincidence analysis might allow for even better background reduction.

\section{Summary}

The newly planned double--$\beta$ decay experiment COBRA consists of a large amount of CZT semiconductor detectors. For such experiments a low background rate in the peak region and a good energy resolution are crucial ingredients. To further improve this approach, CZT semiconductor detectors have been installed and operated in a low background environment deep underground.

An 18\,kg\,day data set was collected with up to 12 1\,cm$^3$ crystals running simultaneously. It was analyzed to determine half life limits on a number of neutrinoless double--$\beta$ decay modes for seven different isotopes. Nearly all of the derived limits improved by roughly an order of magnitude compared to previous results \cite{Bloxham:2007aaa}.

\section{Acknowledgments}

This research was supported by PPARC and the Deutsche Forschungsgemeinschaft (DFG). We would like to thank V. Tretyak for providing the Decay0 code and eV Products for their support. In addition we thank the Forschungszentrum Karlsruhe, especially K. Eitel, for providing the neutron shield and B. Morgan and Y. Ramachers for software. We thank the mechanical workshop of the TU Dortmund for their support and LNGS for giving us the possibility to perform these measurements underground. The work has been supported by the TA-DUSL activity of the ILIAS program (Contract. No. RII3-CT-2004-506222) as part of the EU FP6 programme.
\clearpage
\bibliography{redpaint-16-0nbb}

\end{document}